\documentclass[prb,12pt,superscriptaddress]{revtex4-1}

\usepackage{pdfpages}

\usepackage{amsmath}    
\usepackage{graphicx}   
\usepackage{gensymb}
\usepackage{fancyhdr}
\usepackage{datetime}

\begin{document}

\title{A high-mobility electronic system at an electrolyte-gated oxide surface}
\author{Patrick Gallagher}
\author{Menyoung Lee}
\author{Trevor A. Petach}
\affiliation{Department of Physics, Stanford University, Stanford, California 94305, USA}
\author{Sam W. Stanwyck}
\affiliation{Department of Applied Physics, Stanford University, Stanford, California 94305, USA}
\author{James R. Williams}
\affiliation{Department of Physics, Stanford University, Stanford, California 94305, USA}
\author{Kenji Watanabe}
\author{Takashi Taniguchi}
\affiliation{Advanced Materials Laboratory, National Institute for Materials Science, 1-1 Namiki, Tsukuba, 305-0044, Japan}
\author{David Goldhaber-Gordon}
\affiliation{Department of Physics, Stanford University, Stanford, California 94305, USA}

\maketitle

\textbf{Electrolyte gating is a powerful technique for accumulating large carrier densities in surface two-dimensional electron systems (2DES)~\cite{Fujimoto2013}. Yet this approach suffers from significant sources of disorder: electrochemical reactions can damage or alter the surface of interest~\cite{Schladt2013,Jeong2013,Li2013,Petach2014}, and the ions of the electrolyte and various dissolved contaminants sit Angstroms from the 2DES. Accordingly, electrolyte gating is well-suited to studies of superconductivity~\cite{Ueno2008,Ueno2011,Ye2012} and other phenomena robust to disorder, but of limited use when reactions or disorder must be avoided. Here we demonstrate that these limitations can be overcome by protecting the sample with a chemically inert, atomically smooth sheet of hexagonal boron nitride (BN). We illustrate our technique with electrolyte-gated strontium titanate, whose mobility improves more than tenfold when protected with BN. We find this improvement even for our thinnest BN, of measured thickness 6 \AA, with which we can accumulate electron densities nearing $10^{14}$ cm$^{-2}$. Our technique is portable to other materials, and should enable future studies where high carrier density modulation is required but electrochemical reactions and surface disorder must be minimized.}

A conventional field-effect transistor is controlled by the voltage on a metal electrode separated from the channel by a thin insulating dielectric. The maximum applied voltage is determined by the dielectric breakdown field, beyond which the resistance of the dielectric sharply drops, shorting the metal electrode to the channel. For a typical high-quality dielectric, the breakdown field limits the accumulated carrier density to $\sim 10^{13}$ cm$^{-2}$ (Ref. \citenum{Ahn2003}), although for special cases such as ferroelectrics stronger modulation is possible~\cite{Ahn1999,Takahashi2006,Boucherit2014}. Electrolyte gating circumvents dielectric breakdown by eliminating the metal/dielectric interface: an electrolyte is applied directly to the surface of interest and polarized, drawing one charged species to the surface and building a large electric field. Carrier densities $\sim 10^{15}$ cm$^{-2}$ can be induced by electrolyte gating~\cite{Yuan2009}, facilitating the discovery of superconductivity in new parameter regimes~\cite{Ueno2011,Ye2012} and the creation of novel photonic devices~\cite{Zhang2014}, among other advances.

While very effective at modulating surface properties, electrolyte gating also introduces disorder. The deposition of contaminants on the sample is difficult to control, a problem that is compounded by the possibility of surface-degrading electrochemical reactions. Recent studies have further suggested that chemical modification of the surface of interest, rather than electrostatics, is primarily responsible for the dramatic changes in electronic properties in some electrolyte-gated systems~\cite{Schladt2013,Jeong2013,Petach2014}. Motivated by these challenges, we consider the well-studied 2DES created by electrolyte gating at the surface of strontium titanate (STO)~\cite{Ueno2008,Ueno2010,YLee2011,MLee2011,Li2012,Stanwyck2013,Ueno2014}. The transport properties of this surface 2DES closely resemble those of the 2DES at the lanthanum aluminate/strontium titanate (LAO/STO) interface. However, the highest reported low-temperature electron mobility in the STO 2DES is about 1000 cm$^2$V$^{-1}$s$^{-1}$, at an electron density of $3 \times 10^{13}$ cm$^{-2}$ (Ref. \citenum{Ueno2008}, \citenum{Ueno2010}, \citenum{MLee2011}, \citenum{Ueno2014}); for the same density, the LAO/STO 2DES has mobility up to $10000$ cm$^2$V$^{-1}$s$^{-1}$ (Ref. \citenum{Huijben2013}). We demonstrate that by protecting the STO channel with a thin boron nitride dielectric, the mobility of the resulting electrolyte-gated 2DES substantially increases over a wide density range, surpassing $12000$ cm$^2$V$^{-1}$s$^{-1}$ at a density of $4 \times 10^{13}$ cm$^{-2}$ in our best sample. 

\begin{figure}
\centering
\includegraphics[width=3in]{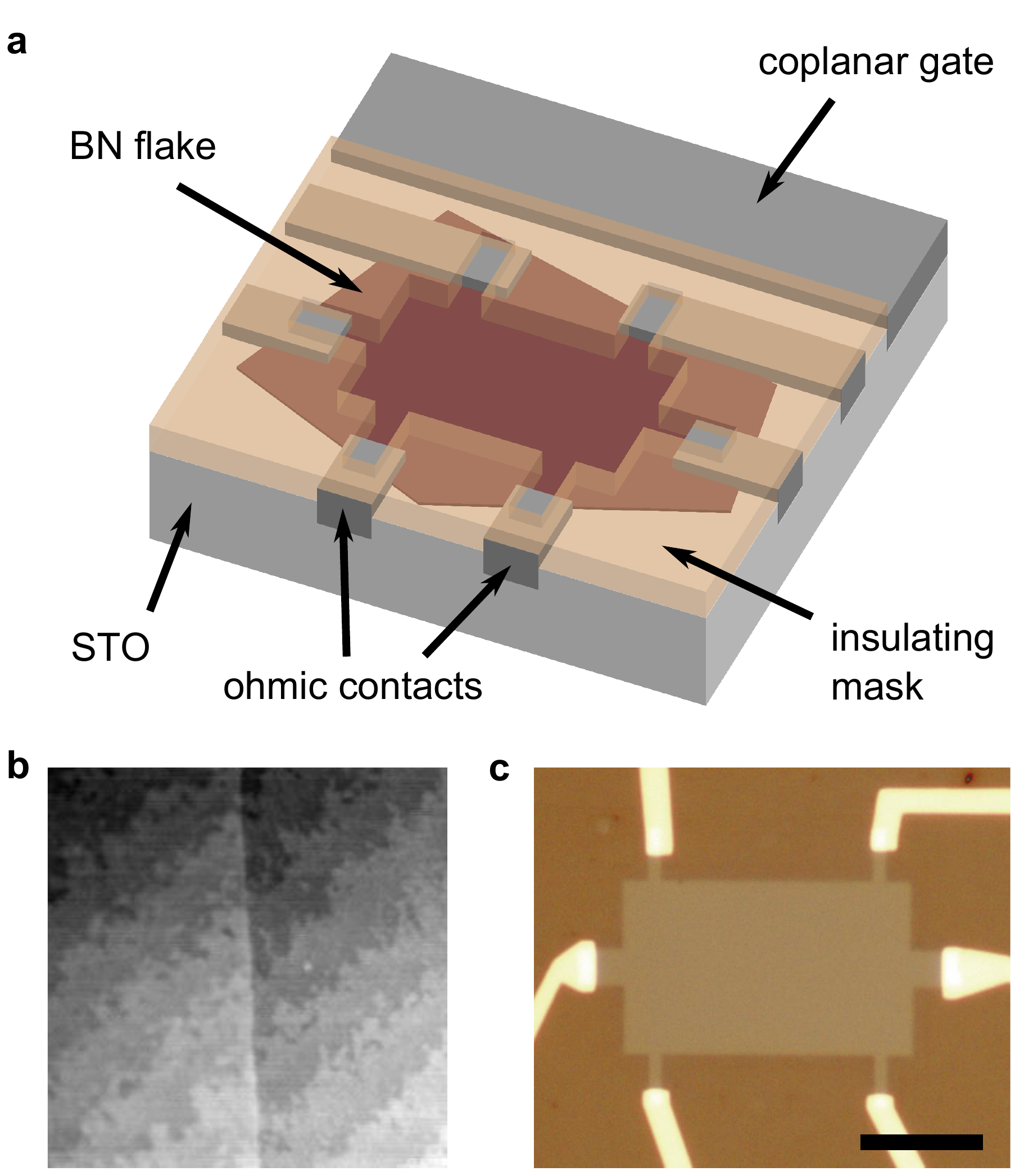}
\caption{
\label{fig:1}
Electrolyte gating with a BN barrier. \textbf{a}, Schematic of a device fabricated on a single crystal of STO. In operation, the entire device is submerged in ionic liquid (not shown), which is polarized by the coplanar gate. \textbf{b}, Atomic force micrograph (topography) of a few-layer BN flake (left half of image) on an STO crystal. STO terraces (4 \AA\ steps) run bottom left to top right, and are visible beneath the BN, indicating that the flake conforms to the substrate with few trapped impurities. Scan window is 1 $\mu$m by 1 $\mu$m. \textbf{c}, Optical micrograph of Sample A, which has a crosslinked PMMA mask (darker brown regions; the relative lightness here is opposite to that in \textbf{a}, where to aid visualization the flake is darker than the mask). The thin BN flake is not visible on STO, but covers the entire opening in the PMMA mask, except near the contacts. Scale bar: 10 $\mu$m.}
\end{figure}

Each of our samples consists of a single crystal of STO partially covered by an atomically-flat BN flake (Fig. \ref{fig:1}a). The BN flake conforms to the substrate without trapping contaminants, as evidenced by the 4 \AA\ terrace steps of the underlying STO seen in the topography of the BN (Fig. \ref{fig:1}b). The substrate is masked by a thick insulator except in a Hall bar-shaped channel area (Fig. \ref{fig:1}c); the electrolyte induces negligible carrier density in the masked regions. In this work, we consider four BN-covered STO samples--denoted A, B, C, and D--with BN thicknesses measured to be 0.6, 1.0, 1.2, and 1.5 nm, respectively, by atomic force microscopy (see Supplementary Information for lateral dimensions and thickness measurement details). For each sample, we collect low-temperature magnetotransport data over multiple cooldowns at different coplanar gate voltages $V_{\rm gate}$.

\begin{figure}
\centering
\includegraphics[width=6in]{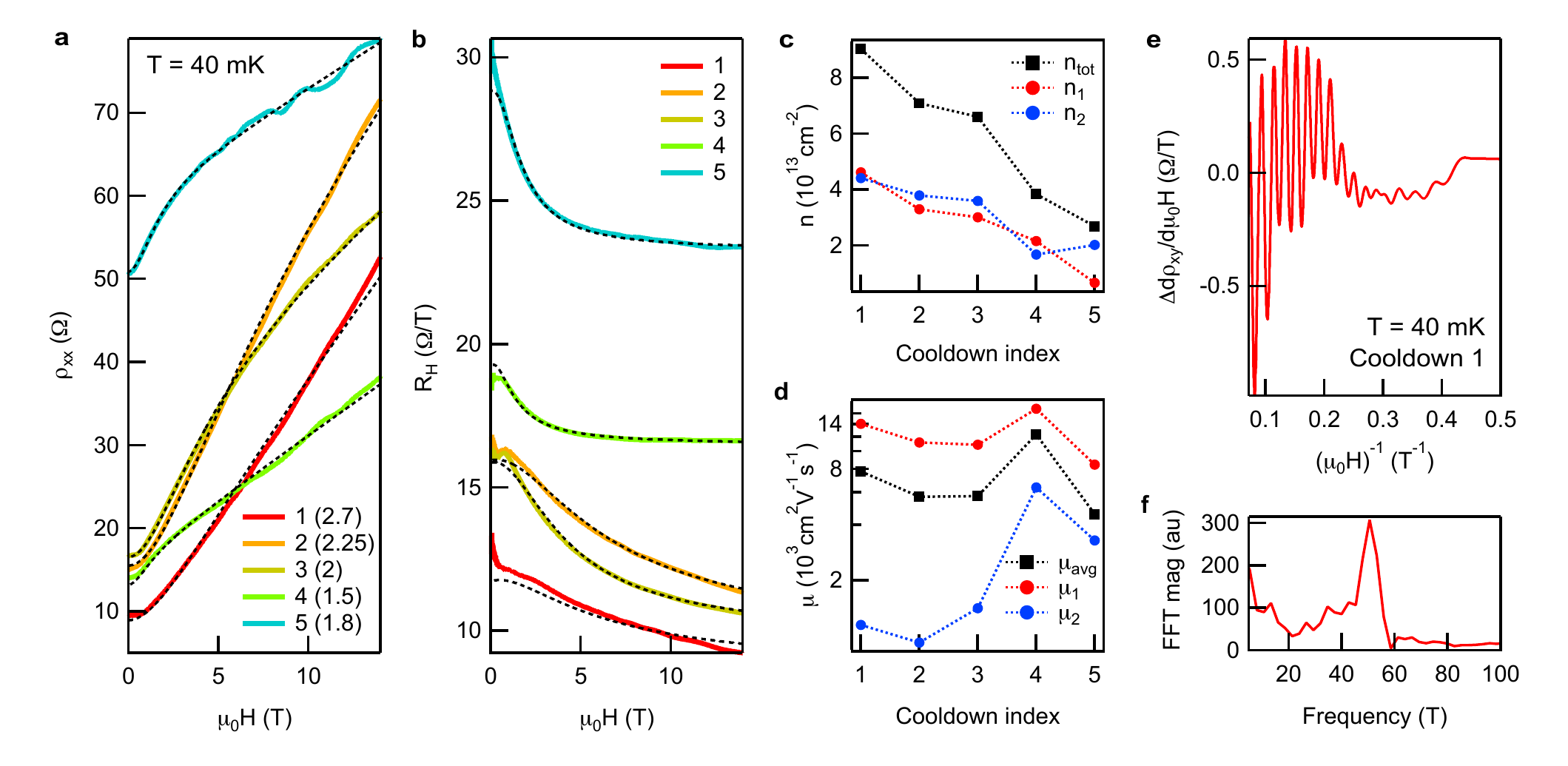}
\caption{
\label{fig:2}
High-mobility magnetotransport in Sample A over a range of densities. \textbf{a}, Sheet resistance $\rho_{xx}$ (symmetrized in field) for 5 cooldowns, labeled 1, 2, 3, 4, and 5, with $V_{\rm gate} = 2.7$ V, 2.25 V, 2 V, 1.5 V, and 1.8 V, respectively. Dashed black curves result from two-band fits with linear magnetoresistance, performed simultaneously on the data in \textbf{a} and \textbf{b}. The BN is 6 \AA\ thick. \textbf{b}, Hall coefficient $R_{\rm H} \equiv \rho_{xy}/\mu_0H$ (symmetrized in field) for the same 5 cooldowns as in \textbf{a} and fits (dashed black curves). \textbf{c}, Extracted carrier densities $n_1$ and $n_2$ for the two bands for each cooldown; $n_{\rm tot} = n_1 + n_2$. \textbf{d}, Extracted carrier mobilities $\mu_1$ and $\mu_2$ for the two bands for each cooldown; $\mu_{\rm avg} = (n_1\mu_1 + n_2\mu_2)/n_{\rm tot}$. \textbf{e}, Quantum oscillations in $d\rho_{xy}/d\mu_0H$ as a function of inverse applied magnetic field for Cooldown 1. For clarity, the signal has been smoothed and a quadratic background has been subtracted. The oscillations commence at $\sim 3$ T for all cooldowns. \textbf{f}, Magnitude of the Fourier transform of \textbf{e}, showing a peak at $\sim 50$ T, corresponding to a density $\sim 2 \times 10^{12}$ cm$^{-2}$.}
\end{figure}

The striking improvement in 2DES quality with a BN spacer is evident in the magnetotransport properties of Sample A, which is covered by a 6 \AA-thick flake (Fig. \ref{fig:2}a,b). The five cooldowns of Sample A, numbered 1 through 5, correspond to different $V_{\rm gate}$ settings. Although higher $V_{\rm gate}$ typically induces higher density, this is not always the case because of hysteresis (see Methods) and because of drifting offset voltages from electrochemical reactions at the gate electrode. To extract density and mobility, we perform a simultaneous fit to the sheet resistance $\rho_{xx}$ and the Hall coefficient $R_{\rm H} \equiv \rho_{xy}/\mu_0H$, where $\rho_{xy}$ is the Hall resistance, $\mu_0$ is the magnetic constant, and $H$ is the applied magnetic field. As is typical in the STO 2DES literature, we assume that the magnetotransport behavior can be described by two bands~\cite{Joshua2012}. Although quantum oscillation data suggest several bands (discussed below), a two-band description often fits the data, providing reliable numbers for average mobility and total density (Supplementary Information). For LAO/STO, a two-band fit with four parameters (densities $n_1$, $n_2$, mobilities $\mu_1$, $\mu_2$) captures the approximate shapes of $\rho_{xx}$ and $R_{\rm H}$, but deviates from the data at higher fields~\cite{Joshua2012}. We encounter the same difficulty: the two-band model cannot simultaneously fit the nonsaturating linear magnetoresistance and nearly-saturated Hall coefficient observed up to 31 T in our samples (Supplementary Information) and in LAO/STO samples~\cite{BenShalom2010}. Inclusion of a third band cannot generally reproduce our high-field data, and where a three-band fit does work, the required densities are unrealistically large, frequently exceeding $10^{16}$ cm$^{-2}$ with mobility $\sim 1$ cm$^2$V$^{-1}$s$^{-1}$. We instead fit to a two-band model in which the sheet resistance of each band contains a term linear in applied field: $\rho_{xx,i} = 1/n_ie\mu_i + k_iH$ for $i = 1,2$ and $k_i \ge 0$. The linear term could arise from spatial fluctuations in mobility~\cite{Parish2003,Kozlova2012}. 

Our two-band fits with linear magnetoresistance provide an excellent match to the data (Fig. \ref{fig:2}a,b). These fits exclude the low-field region, where the magnetotransport properties are affected by magnetic moments in the STO (Ref. \citenum{Joshua2013} and Supplementary Information). We find a high-mobility band with density $n_1$ between $6 \times 10^{12}$ and $5 \times 10^{13}$ cm$^{-2}$ (Fig. \ref{fig:2}c) and mobility $\mu_1$ between 8000 and 17000 cm$^2$V$^{-1}$s$^{-1}$ (Fig. \ref{fig:2}d), as well as a low-mobility band with a similar density $n_2$ and mobility $\mu_2$ that grows with decreasing $n_2$. The total induced density $n_{\rm tot}$ can reach $9 \times 10^{13}$ cm$^{-2}$ (Fig. \ref{fig:2}c) with an average mobility $\mu_{\rm avg} = (n_1\mu_1 + n_2\mu_2)/n_{\rm tot}$ approaching 8000 cm$^2$V$^{-1}$s$^{-1}$ (Fig. \ref{fig:2}d). The average mobility for Cooldown 4 exceeds 12000 cm$^2$V$^{-1}$s$^{-1}$. These mobilities match (for lower densities) and exceed (for higher densities) the highest reported mobilities in LAO/STO 2DES~\cite{Huijben2013,Xie2013}, and are ten times larger than the mobilities reported in the literature for electrolyte-gated STO 2DES at any carrier density~\cite{Ueno2008,Ueno2010,Ueno2014}. Our conclusions are unchanged if we instead calculate $\mu$ and $n$ by naively dividing $R_{\rm H}$ by $\rho_{xx}$, or if we fit with the four-parameter, two-band model (Supplementary Information). 

Quantum oscillations appear above $\sim 3$ T in both $\rho_{xx}$ and $\rho_{xy}$ for all cooldowns. The $\rho_{xy}$ oscillations from Cooldown 1 (Fig. \ref{fig:2}e) show a primary oscillation frequency of 50 T (Fig. \ref{fig:2}f), corresponding to a carrier density near $2 \times 10^{12}$ cm$^{-2}$. This contrasts with the results of the two-band Hall transport fits, in which both bands are at least ten times more populated. For a typical cooldown, we can identify multiple quantum oscillation frequencies corresponding to densities $\sim 10^{12}$ cm$^{-2}$, regardless of the total density measured by the Hall effect. The strongest oscillations thus appear for the lowest Hall densities (see Cooldown 5 in Fig. \ref{fig:2}a), as the bands that produce quantum oscillations now constitute a substantial fraction of the carriers. Our findings resemble quantum oscillation data collected on the highest-mobility LAO/STO 2DES, in which multiple bands of density $\sim 10^{12}$ cm$^{-2}$ show quantum oscillations, and total Hall densities $\sim 10^{13}$ cm$^{-2}$ or lower are required for strong oscillations in $\rho_{xx}$ (Ref. \citenum{McCollam2014}, \citenum{Xie2014}). The presence of low-density oscillating bands does not strongly impact the shapes of $\rho_{xx}$ and $R_{\rm H}$, so the two-band model still captures most of the device behavior (Supplementary Information).

\begin{figure}
\centering
\includegraphics[width=3.8in]{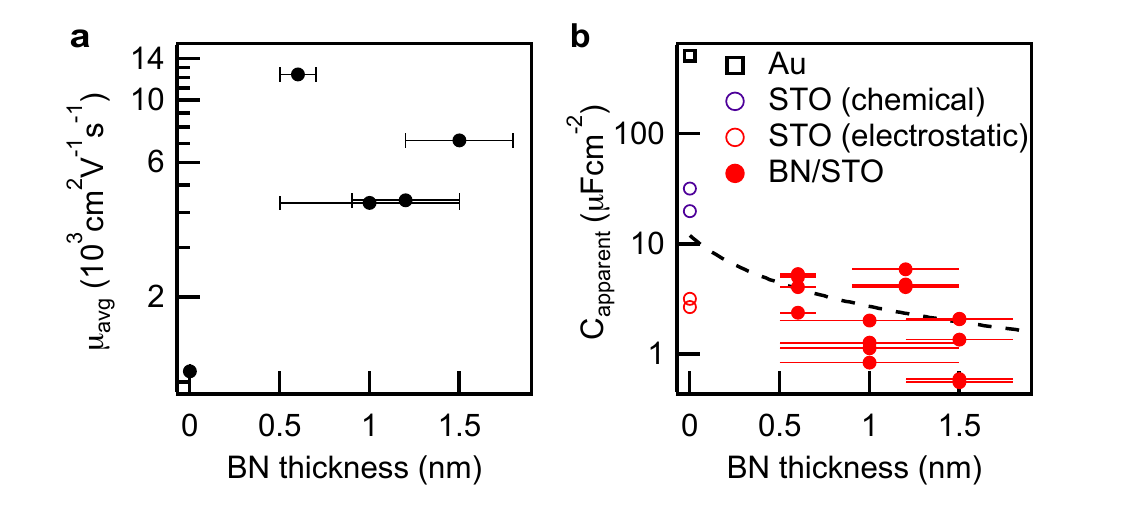}
\caption{
\label{fig:3}
Properties of all measured samples. \textbf{a}, Best average mobility $\mu_{\rm avg}$, extracted from two-band fits with linear magnetoresistance, recorded over all cooldowns at the various BN thicknesses studied. We have included our highest-mobility uncovered STO sample (BN thickness zero). Error bars indicate thickness uncertainty in our atomic force microscope measurements (Supplementary Information). \textbf{b}, Apparent capacitance $C_{\rm apparent} = en_{\rm tot}/V_{\rm gate}$ versus BN thickness for all cooldowns on all BN-covered STO samples (filled red circles). The total density $n_{\rm tot}$ is extracted from two-band fits with linear magnetoresistance. Dashed black line is the maximum capacitance $C_{\rm max}$ for electrostatic carrier accumulation. For comparison, we include $C_{\rm apparent}$ for the bare STO samples from Ref. \citenum{Ueno2010}; those which were determined to be chemically modified (open purple circles) fall above $C_{\rm max}$, while those modulated primarily by electrostatics (open red circles) fall below $C_{\rm max}$. We also show $C_{\rm apparent}$ for the uncovered gold sample from Ref. \citenum{Petach2014} (open black square), which falls far above $C_{\rm max}$. }
\end{figure}

The maximum mobility that we have achieved in each of our four BN-covered samples is significantly higher than the maximum mobility that we have achieved in any uncovered STO sample (Fig. \ref{fig:3}a). The mobility improvement with BN results in part from the added separation between the 2DES and the disordered charges in the electrolyte. As discussed below, we also expect that the BN acts as a barrier to surface-degrading chemical reactions that occur during electrolyte gating or during processing. Our limited sample size produces enough scatter in the maximum mobility as a function of thickness that we cannot identify the main sources of residual disorder.

A single layer of graphene is known to be permeable to atomic hydrogen~\cite{Waqar2007} but impermeable to other small chemical species, including He atoms~\cite{Bunch2008} and Li$^+$ ions~\cite{Das2013}. Because BN has a lattice structure nearly identical to that of graphene, we anticipate a similar diffusion resistance for even our thinnest BN barriers. The energy barrier to diffusion is so high (Ref. \citenum{Das2013} calculates 10 eV for Li$^+$ across graphene) that we still expect impermeability with $V_{\rm gate}$ dropped across our BN. An electrolyte-gated gold sample covered by 6 nm of BN behaved in accordance with these expectations: a gold oxide film is readily grown on uncovered gold samples~\cite{Petach2014}, but the BN-covered gold sample was unmodified (Supplementary Information). The chemical species responsible for the redox reaction is unknown, but these results nonetheless illustrate that BN can limit chemical reactions during electrolyte gating. 

An intriguing possibility for electrolyte-gated oxides is that BN barriers could prevent oxygen removal. Experiments on rutile TiO$_2$ single crystals~\cite{Schladt2013} and VO$_2$ thin films~\cite{Jeong2013} have found evidence that oxygen near the crystal surface diffuses out through the electrolyte, calling into question the relative roles of oxygen vacancy creation and electrostatic carrier accumulation in tuning sample properties. An electrolyte-gating study of STO found that injecting oxygen gas into the electrolyte suppresses the source-drain current, which was interpreted as evidence that the otherwise-observed carrier accumulation results from oxygen vacancies~\cite{Li2013}. Another study of STO concluded that very high gate voltages are required to create oxygen vacancies, and that the reduced STO system (density $\sim10^{15}$ cm$^{-2}$) is three-dimensional and remains conductive at zero gate voltage~\cite{Ueno2010}. While we cannot directly prove the absence of oxygen migration when gating BN-protected STO, we verify that electrostatic carrier accumulation can account for our data by considering the apparent capacitance between the electrolyte and the 2DES, defined as $C_{\rm apparent} = en_{\rm tot}/V_{\rm gate}$. If electrostatics alone is responsible for the carrier accumulation, $C_{\rm apparent}$ should fall below a serial arrangement of two capacitances: that of the double layer formed by the ions ($12\ \mu\text{Fcm}^{-2}$, Ref. \citenum{Ohno2011}), and that of the BN dielectric. This yields $C_{\rm max}^{-1} = (12\ \mu\text{Fcm}^{-2})^{-1} + (4\epsilon_0/t)^{-1}$, where 4 is the dielectric constant of BN and $t$ is the sheet thickness. On the other hand, if carriers accumulate by chemical modification, $C_{\rm apparent}$ is unrestricted. 

For all BN-covered samples, $C_{\rm apparent}$ falls near or below $C_{\rm max}$, and two orders of magnitude below $C_{\rm apparent}$ for uncovered gold, whose surface is chemically modified by electrolyte gating~\cite{Petach2014}. The capacitance to the channel from the large coplanar gate, located less than 200 $\mu$m away, accounts for the violation of the electrostatic limit in Samples A (0.6 nm) and C (1.2 nm). Due to the low-temperature dielectric constant of 25000 in STO and the focusing of field lines from the large gate onto the much smaller Hall bar~\cite{Rakhmilevitch2013}, this capacitance can be as large as several $\mu$Fcm$^{-2}$. We have measured such a capacitance on some samples by zeroing the coplanar gate voltage at low temperature. However, modulating the gate voltage at low temperature appears to cause mechanical problems as our ionic liquid droplet unfreezes upon warmup. We therefore did not collect coplanar gate capacitance data for most samples, and cannot quantitatively correct $C_{\rm max}$. 

Our device geometry exposes some area of our contact metal directly to the electrolyte (Fig. \ref{fig:1}a), limiting $V_{\rm gate}$ to about 3 V: above this, chemical reactions readily occur with the contact metal. This limitation in turn limits the maximum thickness of BN that can be used to create a metallic STO 2DEG. The lowest density for which we have measured a conducting state in our STO 2DES is $10^{13}$ cm$^{-2}$, although the mobility edge may be somewhat lower. To accumulate $10^{13}$ cm$^{-2}$ electrostatically requires a minimum capacitance of 0.5 $\mu$Fcm$^{-2}$, or a maximum BN thickness of 7 nm. Although we have not studied such thick BN flakes, we have measured several samples which had wrinkles in the BN several nm tall due to the transfer process (Supplementary Information). When these wrinkles cut fully across the current path between source and drain, the sample never conducted between source and drain. Presumably the area beneath the wrinkles remained insulating, in approximate numerical agreement with the electrostatic accumulation picture.

The BN barrier need not be kept thin if all conductive material can be masked. This is often difficult in insulators, since the electrolyte must create a conductive path between the device channel and metal contacts, unless the insulator can be chemically doped near the contacts. For intrinsically metallic systems, it is straightforward to mask all conductive area (see our BN on gold sample, Supplementary Information). In this case, higher voltages can in principle be applied without chemical reactions, increasing the maximum thickness of BN that can be used for a target electron density, which may have advantages for certain materials. Our technique is easily applied to other systems, and should enable electrolyte gating experiments that require high carrier mobility, high carrier density, and chemical stability of the surface.

\section*{Methods}

Our samples were fabricated on (100) strontium titanate substrates from either Shinkosha Co. (Japan) or Crystec GmbH (Germany); the vendor for each sample is specified in the Supplementary Information. The surfaces of Shinkosha crystals were TiO$_2$-terminated as received. We prepared a nominally TiO$_2$-terminated surface on the Crystec samples by the method described in Ref. \citenum{Connell2012}. A BN flake was transferred onto the STO surface using the water-based process described in Ref. \citenum{Amet2013}, followed by an anneal for 4 hours at 500\degree C in an Ar/O$_2$ atmosphere. An ohmic contact pattern was defined in PMMA via e-beam lithography at 10 kV, after which the sample was milled with Ar ions at 300 V to etch away the exposed BN and about 40 nm of the underlying STO. Ohmic contacts (10 nm titanium, 40 nm gold) were then deposited into the milled trenches by e-beam evaporation. Finally, an insulating mask with holes to expose the Hall bar and the coplanar gate was patterned using 10 kV e-beam lithography. The mask material was either crosslinked PMMA or sputtered alumina; in both cases a negative process was used so that the channel was not exposed to the e-beam. 

Prior to measurement of each sample, we cleaned the sample surface of resist residues by a brief exposure to a remote oxygen plasma. We then covered the Hall bar and coplanar gate with a drop of the ionic liquid 1-ethyl-3-methylimidazolium bis(trifluoromethanesulfonyl)amide (EMI-TFSI) and placed the sample inside the vacuum chamber of our cryostat (either a dilution refrigerator with base temperature 40 mK or a variable-temperature insert reaching 350 mK or 1.5 K). We polarized the electrolyte at around 290 K, in either high vacuum or helium vapor, by applying a voltage to the coplanar gate. Upon cooling, the polarized electrolyte froze, and we collected magnetotransport data up to the highest available fields (9 T, 14 T, or 31 T) via standard lock-in techniques in a current-biased configuration. We typically used an ac source current of 2 $\mu$A, which exceeded the superconducting critical current in all samples, suppressing the superconducting features that would otherwise appear for some cooldowns in Fig. \ref{fig:2}. The sample was then warmed to near room temperature, melting the electrolyte. We always set $V_{\rm gate}$ to zero in between cooldowns, which introduces some hysteresis in $V_{\rm gate}$.

\section*{Acknowledgements}

We thank Thomas Schladt and Tanja Graf for helpful discussions at an early stage of this work, and Harold Hwang for a careful reading of our manuscript. Sample fabrication was supported by the Air Force Office of Science Research, Award No. FA9550-12-1-02520. Sample measurement was supported by the MURI program of the Army Research Office, Grant No. W911-NF-09-1-0398. Development of the ionic liquid gating technique was supported by the Center on Nanostructuring for Efficient Energy Conversion (CNEEC) at Stanford University, an Energy Frontier Research Center funded by the U.S. Department of Energy, Office of Basic Energy Sciences under Award No. DE-SC0001060. P.G. acknowledges support from the DOE Office of Science Graduate Fellowship Program. M.L. acknowledges support from Stanford University. J.R.W. and D.G.-G. acknowledge support from the W. M. Keck Foundation. 

A portion of our sample fabrication and characterization was performed at the Stanford Nano Center (SNC)/Stanford Nanocharacterization Laboratory (SNL), part of the Stanford Nano Shared Facilities. A portion of our measurements was performed at the National High Magnetic Field Laboratory, which is supported by National Science Foundation Cooperative Agreement No. DMR-1157490, the State of Florida, and the U.S. Department of Energy.

\section*{Author contributions}

P.G., J.R.W., and D.G.-G. designed the experiment. P.G. fabricated the BN on STO samples and performed the measurements, with help from M.L., S.W.S., and J.R.W. P.G., M.L., and D.G.-G. analyzed the data. T.A.P. performed the BN on gold experiment. K.W. and T.T. grew the BN crystals. P.G. prepared the manuscript with input from all authors. 

\section*{Competing financial interests}

The authors declare no competing financial interests.

\newpage



\section*{Supplementary material}

# CONTENTS

| Sample name | Crystal vendor | Mask material | BN thickness | $L_{\text{tot}}$ | $L_c$ | $W$ | $D_{\text{coplanar}}$ |
|---|---|---|---|---|---|---|---|
| Sample A | Shinkosha | PMMA | $0.6 \pm 0.1$ nm | 23 $\mu$m | 18.5 $\mu$m | 14 $\mu$m | 126 $\mu$m |
| Sample B | Crystec | alumina | $1.0 \pm 0.5$ nm | 9 $\mu$m | 6 $\mu$m | 7.5 $\mu$m | 84 $\mu$m |
| Sample C | Shinkosha | PMMA | $1.2 \pm 0.3$ nm | 6 $\mu$m | 4 $\mu$m | 4 $\mu$m | 194 $\mu$m |
| Sample D | Crystec | alumina | $1.5 \pm 0.3$ nm | 10 $\mu$m | 7 $\mu$m | 6 $\mu$m | 125 $\mu$m |

TABLE S1. Sample details. $L_{\text{tot}}$ is the total length of the Hall bar channel, and $W$ is the channel width. $L_c$ is the center-to-center distance between $R_{xx}$ voltage probes. $D_{\text{coplanar}}$ is the distance from the Hall bar to the nearest point on the coplanar gate.

## I. SAMPLE PROPERTIES

#### a. Lateral dimensions

The samples studied in this work have different lateral dimensions because we are constrained by the dimensions of the chosen BN barrier flake. We are also typically forced to position our coplanar gate and ohmic contacts away from thick, randomly-positioned flakes that are unintentionally transferred alongside the flake of interest, leading to further geometry variations. Nonetheless, the approximate geometry in all cases is as shown in Fig. 1a in the main text, with a small Hall bar about 100 $\mu$m from a coplanar gate. The gate dimensions are on the order of 1 mm by 0.2 mm: the gate must be much larger than the Hall bar so that the applied voltage drops across the BN, instead of at the double layer at the coplanar gate. Ohmic traces are a few microns wide and extend to bond pads about 1 mm away from the Hall bar and coplanar gate. Some specific sample dimensions are listed in Table S1.

#### b. Flake thickness

We determined the flake thicknesses reported in Table S1 by atomic force microscopy (AFM) after transferring the flakes onto the STO crystal and annealing at 500°C. We found that after annealing, a low density of small, mobile particles remained on the sample surface. While these particles were too small and too sparse to affect transport measurements, they could easily be collected by the AFM tip and redeposited, particularly at the step edge

3

whose height we wanted to measure. This introduced sizable errors in our tapping mode AFM (TAFM) measurements. The thickness uncertainties quoted in Table S1 reflect the measured thickness variations between linescans at different positions on the same flake. Sample B was particularly difficult to image with TAFM because of mobile particles, and therefore has a large error bar.

For our later generation of samples (Samples A and C), we used contact AFM (CAFM), which is more suitable than TAFM for studying a surface with mobile particles, and generally leads to a more accurate step height measurement between substrate and flake[1,2]. However, the difference in friction between STO and few-layer h-BN causes twisting of the CAFM cantilever and thus a different measured flake thickness for left-moving and right-moving scans[1]. Because the offset between left-moving and right-moving scans is small (several Å), to first order the correct step height can be found by averaging the step heights measured in each direction. This average thickness is the thickness quoted in Table S1 for Samples A and C, and the quoted uncertainties reflect the variation in average thickness between linescans at different positions on the same flake. As a sanity check, we also performed TAFM on Samples A and C, and found thicknesses consistent with the average thickness measurements from CAFM.

### c. Determination of $\rho_{xx}$ and $\rho_{xy}$

In this study, the quantities that we directly measure as a function of magnetic field are the longitudinal resistance $R_{xx}(H)$ and the transverse resistance $R_{xy}(H)$ of our six-terminal devices. To extract the sheet resistance $\rho_{xx}(H)$ of the 2DES from the measured signal $R_{xx}(H)$, $R_{xx}(H)$ must be multiplied by a factor that accounts for geometrical details such as the number of squares between the voltage probes, the nonzero width of the voltage probes, and the distance between the voltage probes and the current leads. Similarly, to determine the Hall resistance $\rho_{xy}(H)$, the measured transverse resistance $R_{xy}(H)$ must also be corrected for a geometrical factor, since in some geometries the Hall voltage can be shorted out by the current leads[3]. In general, these geometrical correction factors may depend on the Hall angle $\theta_\mathrm{H} = \tan^{-1}(\rho_{xy}(H)/\rho_{xx}(H))$.

We have used the conformal mapping technique[4] to numerically determine the geometrical correction factors $C_{xx} \equiv R_{xx}/\rho_{xx}$ and $C_{xy} \equiv R_{xy}/\rho_{xy}$ for the geometry of each sample.

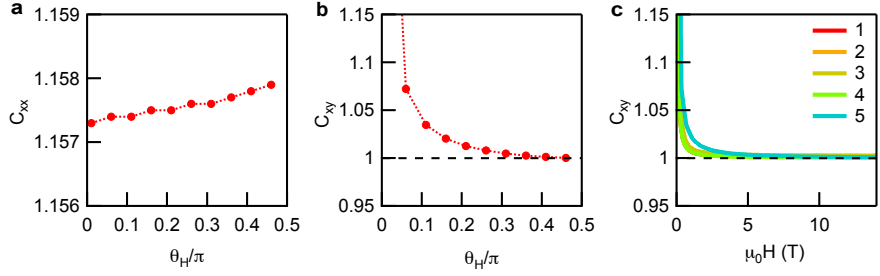

FIG. S1. Conformal mapping results for Sample A. **a**, Calculated ratio $C_{xx}$ of longitudinal resistance to sheet resistance for Hall angles $\theta_\text{H}$ between $0.01\pi$ and $0.46\pi$. **b**, Calculated ratio $C_{xy}$ of transverse resistance to Hall resistance. The black dashed line shows the expected ratio $C_{xy} = 1$ for an ideal Hall bar. Numerical errors in the conformal mapping are most significant near $\theta_\text{H} = 0$, and are likely responsible for the deviation of $C_{xy}$ from 1. **c**, The calculation in **b**, interpolated and plotted as a function of applied field for the five cooldowns of Sample A (Fig. 2). The cooldown index for each curve is indicated in the legend.

Conformal mapping calculations were performed using the Schwarz-Christoffel Toolbox for MATLAB, created by Tobin Driscoll (http://www.math.udel.edu/~driscoll/SC/index.html). For all samples and all experimentally relevant Hall angles $\theta_\text{H}$, we find that $C_{xy} \approx 1$, and $C_{xx} \approx L_c/W$, where $L_c$ is the center-to-center distance between longitudinal voltage probes and W is the channel width. We therefore compute $\rho_{xx}(H) = C_{xx} R_{xx}(H)$, where $C_{xx}$ is the field-independent constant calculated by conformal mapping, and we set $\rho_{xy}(H) = R_{xy}(H)$ so that the Hall coefficient $R_\text{H}(H) \equiv \rho_{xy}(H)/\mu_0 H = R_{xy}(H)/\mu_0 H$.

For concreteness, we describe in detail the results of the conformal mapping for Sample A. The calculated longitudinal correction factor $C_{xx}(\theta_\text{H})$ varies by less than 0.1% over the full range of Hall angles (Fig. S1a). The calculated $C_{xy}(\theta_\text{H})$ converges to 1 as $\theta_\text{H}$ approaches $\pi/2$ (Fig. S1b); minor deviations from 1 at lower $\theta_\text{H}$ likely result from numerical errors in the conformal mapping, which grow large for smaller $\theta_\text{H}$. Regardless of the cause of the deviations, most of the experimental field range corresponds to $\theta_\text{H}$ sufficiently large that $C_{xy} = 1$ (Fig. S1c).

## II. FITTING DETAILS

### a. Fitting procedure

All fitting described in this work has been carried out using the Ceres Solver (http://ceres-solver.org/), an open-source C++ library for solving nonlinear least squares problems. Our fitting procedure takes two input datasets, $\rho_{xx}(H)$ and $R_H(H)$, and finds the single set of two-band model parameters that best fits both datasets. We use a least-squares residual in which each term is normalized by the corresponding value of $\rho_{xx}(H)$ or $R_H(H)$, so that the fitting does not favor high-field data over low-field data:

$$\text{Residual} \equiv \sum_H \left( \frac{(\rho_{xx}(H) - f_{xx}(H; n_1, n_2, ...))^2}{\rho_{xx}(H)^2} + C \frac{(R_H(H) - f_H(H; n_1, n_2, ...))^2}{R_H(H)^2} \right),$$

where $f_{xx}$ and $f_H$ are the two-band predictions for $\rho_{xx}$ and $R_H$, and $C$ is a constant that adjusts the relative importance of $R_H$ over $\rho_{xx}$ in the fitting. Typically we set $C = 1$, but for some datasets, in particular those with strong $\rho_{xx}$ oscillations, we increase $C$ to get appropriate fits. The constraints on the fit parameters are very loose: parameters are allowed to range over several orders of magnitude. Our best fits are reproducible over a large range of initial conditions provided to the fitting algorithm.

### b. Linear magnetoresistance model

Our fits presented in the main text use a two-band model in which the resistivity tensor of each band is described by

$$\rho_{xx,i} = \rho_i + k_i \mu_0 H$$

$$\rho_{xy,i} = \mu_0 H / n_i e,$$

where $\rho_i \equiv 1/n_i e \mu_i$ is the zero-field resistivity of band $i$, $n_i$ is the density of band $i$, $k_i \geq 0$, and $\mu_0$ is the magnetic constant (in the main text we absorbed $\mu_0$ into $k_i$ for simplicity). The conductivity tensor of the complete system is the sum of the conductivity tensors of the individual bands.

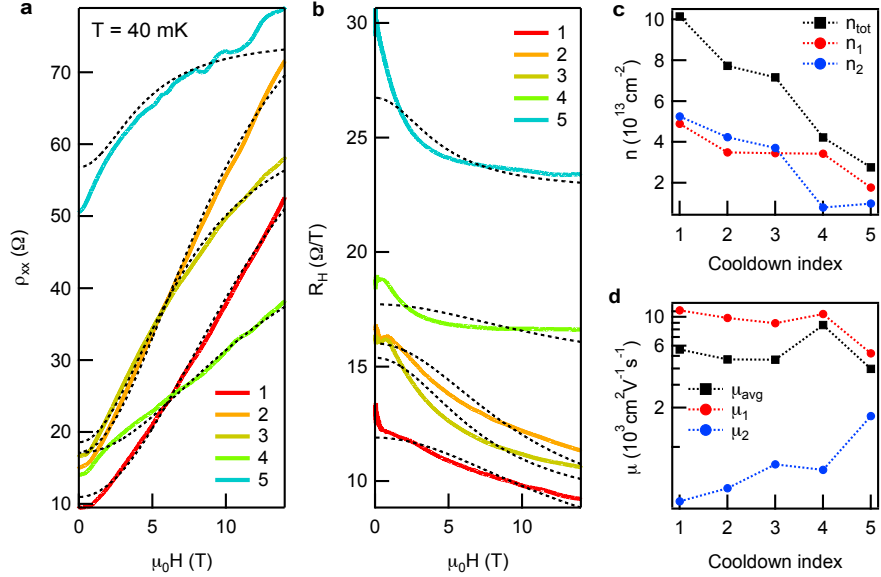

FIG. S2. Four-parameter, two-band fits to the data in Fig. 2. **a**, Sheet resistance $\rho_{xx}$ versus applied magnetic field for the 5 cooldowns in the main text. Black dashed curves are fits to the standard two-band model. **b**, Hall coefficient $R_H$ versus applied magnetic field for the same 5 cooldowns. Fits (black dashed curves) are performed simultaneously on the $\rho_{xx}$ and $R_H$ data for a given cooldown. **c**, Densities $n_1$ and $n_2$ of the two bands; $n_{\text{tot}} = n_1 + n_2$. **d**, Mobilities $\mu_1$ and $\mu_2$; $\mu_{\text{avg}} = (n_1\mu_1 + n_2\mu_2)/n_{\text{tot}}$.

### c. Fitting more than two bands with a two-band model

It is clear from the presence of quantum oscillations with many frequencies that more than two bands contribute to transport. This is also the case for the highest-mobility LAO/STO 2DES, in which several bands contribute to quantum oscillations even for total carrier densities as low as $10^{13}$ cm$^{-2}$ (Ref. 5, 6). Despite the large number of bands in our 2DES, the two-band model with linear magnetoresistance presented in the text captures the essential magnetotransport behavior. In fact, fits to the data in Fig. 2 in the main text are also possible with the four-parameter, two-band model[7], although the fit quality is lower (Fig. S2a,b). The extracted densities (Fig. S2c) and mobilities (Fig. S2d) are still comparable to those shown in the main text.

The two-band model can give a reasonable fit to the many-band data by averaging the set of high mobility bands into one band, and the set of low-mobility bands into another

| Band index $i$ | $n_i$ (cm$^{-2}$) | $\mu_i$ (cm$^2$V$^{-1}$s$^{-1}$) |
|---|---|---|
| 1 | $1\times10^{12}$ | 20000 |
| 2 | $2\times10^{12}$ | 15000 |
| 3 | $2\times10^{12}$ | 12000 |
| 4 | $3\times10^{12}$ | 8000 |
| 5 | $3\times10^{12}$ | 6000 |
| 6 | $3\times10^{12}$ | 5000 |
| 7 | $2\times10^{12}$ | 4000 |
| 8 | $3\times10^{12}$ | 3000 |
| 9 | $4\times10^{12}$ | 2000 |
| 10 | $2\times10^{12}$ | 1000 |

TABLE S2. Parameters for ten bands used to generate data in Fig. S3. These parameters are chosen to be plausible given the densities extracted from quantum oscillations and the average mobility and total density that we extract from the Hall analysis, but are otherwise arbitrary.

band. For instance, if we assume the existence of ten bands with the arbitrarily-chosen (but plausible) densities and mobilities shown in Table S2, the simulated $\rho_{xx}$ and $R_\mathrm{H}$ are acceptably fit by the four-parameter, two-band model (Fig. S3a,b). The resulting fit parameters are $n_1 = 1.1 \times 10^{13}$ cm$^{-2}$, $n_2 = 1.4 \times 10^{13}$ cm$^{-2}$, $\mu_1 = 11267$ cm$^2$V$^{-1}$s$^{-1}$, and $\mu_2 = 2467$ cm$^2$V$^{-1}$s$^{-1}$. These values yield a total density $n_\mathrm{tot} \equiv \sum_i n_i = 2.5 \times 10^{13}$ cm$^{-2}$ and an average mobility is $\mu_\mathrm{avg} \equiv \sum_i n_i \mu_i / n_\mathrm{tot} = 6395$ cm$^2$V$^{-1}$s$^{-1}$, which agree well with the total density $2.5 \times 10^{13}$ cm$^{-2}$ and average mobility 6320 cm$^2$V$^{-1}$s$^{-1}$ calculated from Table S2.

The six-parameter fit with linear magnetoresistance yields better agreement to the $\rho_{xx}$ and $R_\mathrm{H}$ curves, even though no linear magnetoresistance is used to produce the ten-band curves (Fig. S3a,b). The fit parameters are $n_1 = 9.2 \times 10^{12}$ cm$^{-2}$, $n_2 = 1.5 \times 10^{13}$ cm$^{-2}$, $\mu_1 = 12382$ cm$^2$V$^{-1}$s$^{-1}$, $\mu_2 = 3039$ cm$^2$V$^{-1}$s$^{-1}$, $k_1 = 0$, and $k_2 = 1.6$ $\Omega$/T, which give $n_\mathrm{tot} = 2.4 \times 10^{13}$ cm$^{-2}$ and $\mu_\mathrm{avg} = 6587$ cm$^2$V$^{-1}$s$^{-1}$, also in good agreement with the correct values.

On the other hand, if we assume that there are in fact two higher-density bands in our system, the addition of several low-density oscillating bands does not strongly affect the shapes of the magnetotransport curves. For a specific example, consider adding four bands, each of density $10^{12}$ cm$^{-2}$, with mobilities 5000, 6000, 8000, and 12000 cm$^2$V$^{-1}$s$^{-1}$ to the

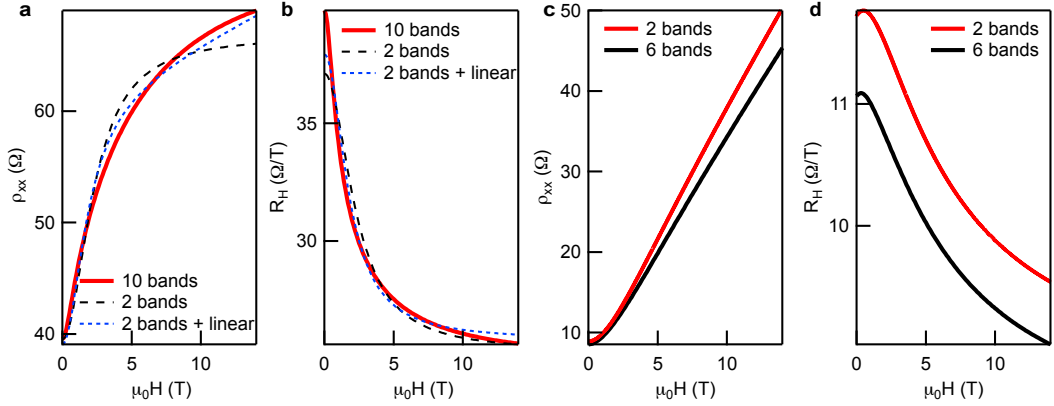

FIG. S3. Capturing many-band data with a two-band model. **a** and **b**, Simulated sheet resistance $\rho_{xx}$ and Hall coefficient $R_H$ versus applied magnetic field for the 10-band system whose parameters are given in Table S2 (red curves). Black dashed curves result from a fit to the four-parameter, two-band model. Blue dashed curves result from a fit to the six-parameter, two-band model with linear magnetoresistance. **c** and **d**, Red curves: six-parameter fits to data from Sample A, Cooldown 1. Black curves: same as red curves, except that four low-density bands have been added to the calculation (see text).

two bands with linear magnetoresistance extracted in the main text (Fig. 2) for Sample A, Cooldown 1. The four additional bands have linear magnetoresistance coefficients of zero, for simplicity. Other than globally shifting $R_H$ because of the increased density, the addition of these bands has a modest effect on $\rho_{xx}$ and $R_H$ (Fig. S3c,d).

We conclude that regardless of the actual number of bands, the two-band model with or without linear magnetoresistance is a reliable method for extracting total density and average mobility.

### d. Justification of two-band fits with linear magnetoresistance

The fact that the six-parameter fit could more closely approximate the ten-band $\rho_{xx}$ and $R_H$ (Fig. S3a,b) might suggest that the linear terms that we have introduced in these fits are not physically meaningful. However, in many of our samples we observe a nonsaturating, approximately linear $\rho_{xx}$ despite nearly saturating $R_H$. Such behavior is particularly evident for Sample A, Cooldown 4; here the data are poorly fit by the four-parameter, two-band

9

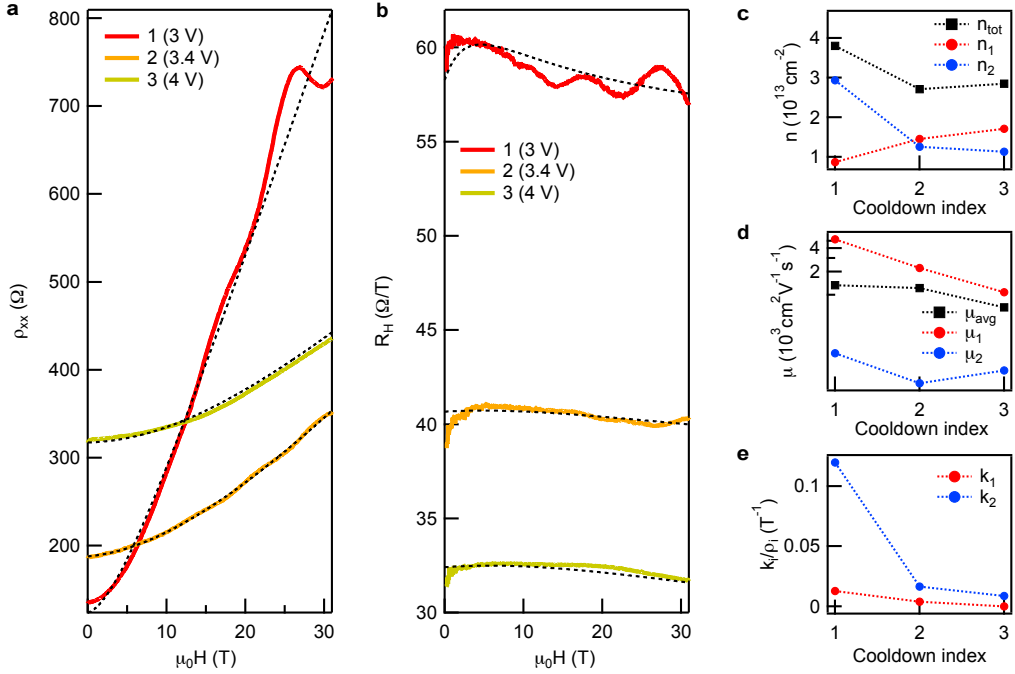

FIG. S4. Linear magnetoresistance measured up to 31 T in Sample B. **a**, Sheet resistance $\rho_{xx}$ versus applied magnetic field for three successive cooldowns, labeled 1, 2, and 3. The gate voltage used for each cooldown is shown in parentheses in the legend. All cooldowns (but especially Cooldown 1) display nonsaturating and approximately linear magnetoresistance up to the maximum field of 31 T. Black dashed curves are fits to our two-band model with linear magnetoresistance. The sample is immersed in liquid $^3$He at 350 mK. **b**, Hall coefficient $R_H$ for the three cooldowns. Fits (black dashed curves) are performed simultaneously on the $\rho_{xx}$ and $R_H$ data for a given cooldown. **c**, Densities $n_1$ and $n_2$ of the two bands; $n_{\text{tot}} = n_1 + n_2$. **d**, Mobilities $\mu_1$ and $\mu_2$; $\mu_{\text{avg}} = (n_1 \mu_1 + n_2 \mu_2)/n_{\text{tot}}$. **e**, Linear magnetoresistance coefficients $k_i$, for $i = 1, 2$, normalized to the zero field resistivities $\rho_i$ of the corresponding bands.

model (Fig. S2), but very well fit by the six-parameter, two-band model (Fig. 2).

The linear term is further justified by our observation in Sample B of nonsaturating linear magnetoresistance despite nearly saturated $R_H$ up to fields as high as 31 T (Fig. S4a) at the National High Magnetic Field Lab. Sample B is the only sample that we measured in the Magnet Lab. Although this sample exhibits lower mobility (Fig. S4d) than other BN-covered STO samples, which we ascribe to subpar cleanliness of our procedures away from our home lab, the qualitative magnetoresistance behavior is similar to that of our

10

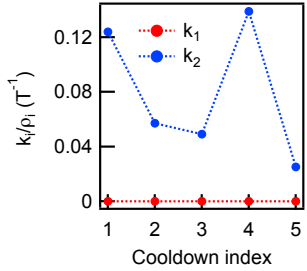

FIG. S5. Linear magnetoresistance coefficients in Fig. 2 fits, normalized to the zero field resistivities $\rho_i$ of the corresponding bands. The fits resulted in a high-mobility band (band 1) with $k_1 = 0$ for all cooldowns.

other samples measured at lower fields. Nonsaturating linear magnetoresistance with an apparently saturating $R_H$ has also been observed in LAO/STO up to 31 T (Ref. 8).

We have attempted to fit our data from all samples with a three-band model (parameters $n_1, n_2, n_3, \mu_1, \mu_2, \mu_3$), since a high-density, low-mobility band could in principle give rise to a nearly linear growth of $\rho_{xx}$ with a nearly constant $R_H$ for the experimentally-accessible field range. However, the three-band model does not fit all of our datasets, while the two-band model with linear magnetoresistance can generally fit the data very well. Even when the three-band fits do capture the data, the resulting densities are surprisingly high, ranging from $10^{14}$ to $10^{16}$ cm$^{-2}$, with mobilities $\sim 1$ cm$^2$V$^{-1}$s$^{-1}$. The two-band model with linear magnetoresistance is our best known method to accurately fit our data.

### e. Interpretation of $k_i$

In our six-parameter, two-band fits with linear magnetoresistance coefficients $k_i$, we typically find that the high-mobility band has a smaller linear coefficient than does the low-mobility band, even when $k_i$ is normalized to the zero-field resistivity $\rho_i$ (Fig. S4e). In fact, our fits in Fig. 2 in the main text all yielded exactly zero for $k_1$, the linear magnetoresistance coefficient of the high-mobility band (Fig. S5). One possible interpretation of this result, assuming that the linear magnetoresistance arises from spatial fluctuations in the mobility[9,10], is that the low-mobility band is nearer the STO surface, while the high-mobility band is buried deeper. The inhomogeneities responsible for the mobility fluctuations plausibly reside on the surface, thereby more strongly affecting the low-mobility band.

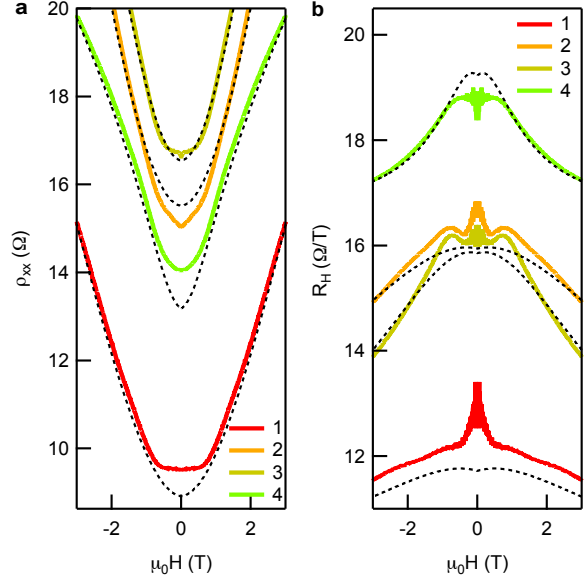

FIG. S6. Low-field data and fits from Fig. 2, for cooldowns 1-4. **a**, Sheet resistance $\rho_{xx}$ and fits (black dashed curves). **b**, Hall coefficient $R_H$ and fits (black dashed curves). The $\rho_{xx}$ and $R_H$ data cannot be simultaneously fit by our model for fields below $\sim 1$ T.

However, we caution that the linear term is only the first-order correction to the band resistivities $\rho_{xx,i}$. The linear magnetoresistance due to mobility fluctuations exhibits a crossover field, below which the magnetoresistance is quadratic[9,10]. Giving the linear magnetoresistance mainly to the low-mobility band may simply help approximate this shape: if the densities of the two bands are similar, the low-mobility band becomes relevant to the magnetoresistance only at higher fields.

### f. Low-field data

All of the fits shown in the main text and in the Supplementary Information neglect the low-field region, where magnetic effects are likely important[11]. We find that the behavior of $\rho_{xx}$ and $R_H$ in this region cannot be simultaneously captured by few-band physics, even with linear magnetoresistance terms (Fig. S6). Cooldowns 1, 2, and 3 had superconducting features, with $T_c = 60$, 115, and 185 mK, respectively. These features do not appear in Fig. S6 because the critical current in each case is much smaller than the 2 $\mu$A ac excitation. When measured with a smaller ac excitation, the superconducting features are suppressed

12

above $H_c \sim 50$ mT.

### III. CHEMICAL RESISTANCE OF BN FLAKES

An electrolyte-gating study of gold films has reported carrier density modulation of up to $3 \times 10^{15}$ cm$^{-2}$ and correspondingly large changes in sheet resistance[12]. The capacitance required for these changes exceeds by an order of magnitude the expected capacitance of the electrolyte gate[13], suggesting that chemical modification of the gold film is responsible for the modulation. Indeed, we have shown that electrolyte gating readily oxidizes the gold film at negative gate voltages, and that this oxide growth accounts for at least 90% of the observed changes in transport properties[14].

To determine whether our BN films can prevent this sort of electrochemical reaction, we transferred a BN flake 6 nm thick onto a gold Hall bar 30 nm thick. The topography scan of the resulting structure shows that the flake mostly conforms to the Hall bar (Fig. S7a). However, because the Hall bar protrudes 30 nm from the oxide substrate, the flake relaxes strain by forming several wrinkles. The flake is also not perfectly flush with the gold where top two voltage probes meet the channel. Prior to electrolyte gating, all unprotected gold areas were masked by a layer of crosslinked PMMA (Fig. S7b). The protected Hall bar and the on-chip coplanar gate were then covered in the ionic liquid N,N-diethyl-N-methyl-N-2-methoxyethyl ammonium tetrafluoroborate (DEME-BF4) and inserted in a vacuum chamber at room temperature.

First, we swept the gate voltage between -2 and 2 V. The total change in $\rho_{xx}$ over these 4 V is $\sim 5 \times 10^{-4}$ $\Omega$ (Fig. S7c). For 4 V dropped across a BN flake 6 nm thick, electrostatics gives a density change $\Delta n = 4\epsilon_0 \times 4$ V/6 nm $= 1.5 \times 10^{13}$ cm$^{-2}$ (we neglect the double layer capacitance, which is at least ten times the capacitance from the 6 nm BN dielectric). The total 2D carrier density in a gold film 30 nm thick is $n = 1.8 \times 10^{17}$ cm$^{-2}$. If the sheet resistance is proportional to $1/n$, electrostatics predicts a fractional change in sheet resistance $\sim 10^{-4}$, close to the fractional change $\sim 3 \times 10^{-4}$ that we observe. Considering that we have neglected changes in electron mobility and the inhomogeneous distribution of the induced carrier density within the gold film, this approximate agreement is consistent with purely electrostatic carrier accumulation. The gate leakage current through the electrolyte and into the sample (Fig. S7d) is well below the 10 pA resolution of our measurement

13

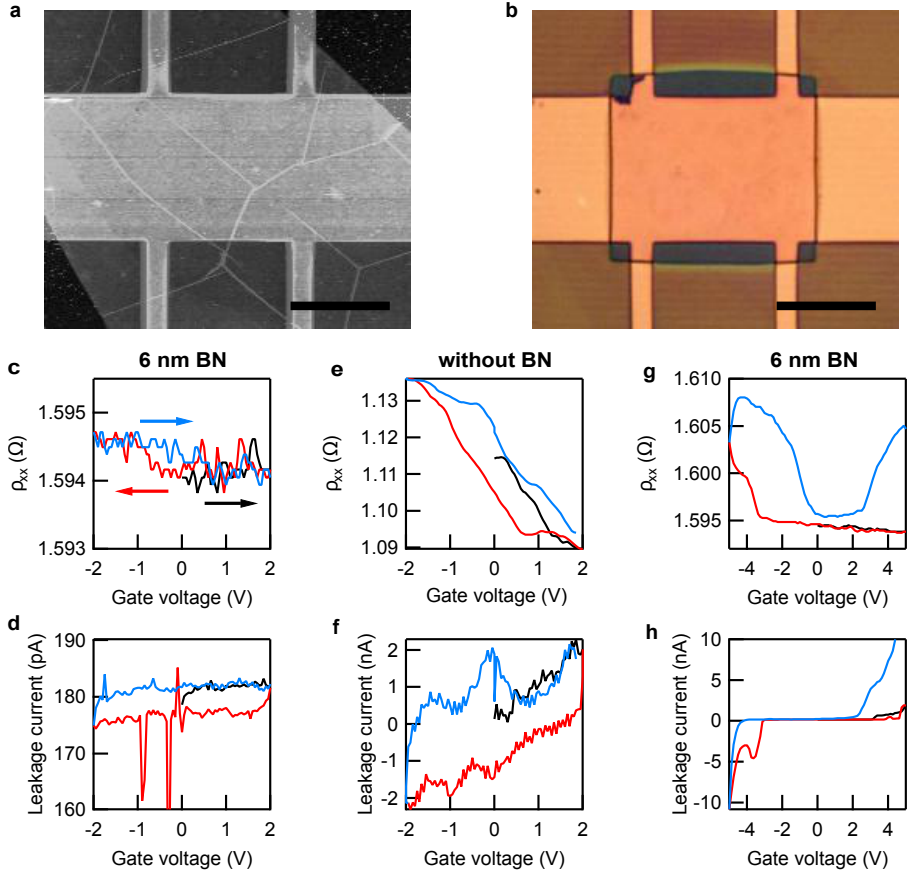

FIG. S7. Electrolyte gating of gold on SiO$_2$ with a BN barrier. **a**, AFM topography of the 30-nm-thick gold Hall bar, covered with a 6-nm-thick BN flake, which runs from top left to bottom right. The flake has visible wrinkles and does not conform perfectly to the Hall bar where the channel meets the top two voltage probes. Scale bar: 10 $\mu$m. **b**, Optical micrograph of the same device after measurement. A PMMA mask was deposited prior to measurement to ensure that no gold was exposed directly to the electrolyte. The flake is faintly visible beneath the PMMA mask. The upper left voltage lead shows damage (see text). Scale bar: 10 $\mu$m. **c**, Sheet resistance $\rho_{xx}$ for a sequence of gate voltage sweeps, in vacuum at 300 K. Black trace: sweeping from 0 to 2 V. Red trace: 2 to -2 V. Blue trace: -2 to 2 V. **d**, Leakage current through the electrolyte for the same sweeps; there is a $\sim$ 180 pA offset in our measurement electronics. The measured leakage current is zero to within our sensitivity. **e** and **f**, Equivalent dataset to **c** and **d** on a gold device of similar thickness but without BN, showing significant leakage current and $\rho_{xx}$ modulation. **e** and **f**, Sheet resistance and leakage current for sweeps between ±5 V, performed after the sweeps in **c** and **d**. The leakage becomes nonzero as the gate reaches 3 V for the first time (black trace); $\rho_{xx}$ begins to change significantly after the gate reaches -3 V.

14

electronics, further implying the absence of electrochemical reactions, which would involve charge transfer from the electrolyte to the sample.

In contrast, electrolyte gating of an unprotected gold Hall bar with similar film thickness shows a very strong modulation of $\rho_{xx}$ and a large leakage current even for very small gate voltages (Fig. S7e,f). The fractional change in $\rho_{xx}$ over the -2 to 2 V window is $\sim 130$ times larger for the uncovered sample than it is for the covered sample. This cannot be fully accounted for by the added thickness of the BN spacer, which should decrease the parallel plate capacitance from the double layer value of 7 $\mu$Fcm$^{-2}$ (Ref. 14) by a factor of 12. Thus, the fractional change in $\rho_{xx}$ for a given electric field is 10 times smaller in the BN-protected sample than it is in the unprotected sample, implying that the BN very strongly suppresses, if not entirely prevents, the oxidation of gold by the electrolyte.

We next proceeded to higher gate voltages on our BN-covered sample. In our first sweep up from 0 to 5 V, the leakage current started to noticeably increase at around 3 V (Fig. S7h), but the change in $\rho_{xx}$ remains small (Fig. S7g). However, upon sweeping from 5 to -5 V, we find a sudden $\rho_{xx}$ increase around -3 V, coincident with a leakage current spike. The next sweep, from -5 to 5 V, shows a significant $\rho_{xx}$ modulation even for positive gate voltages. The fractional magnitude of this modulation per volt is comparable to that of the unprotected gold Hall bar (Fig. S7e).

These data strongly suggest that our BN flake effectively blocks the oxidation reaction at the gold surface. At a sufficiently high voltage (approximately $\pm$ 3 V), the BN barrier itself starts to degrade, allowing reactions with the gold surface. The damage to the BN flake and/or the underlying gold is visible in an optical micrograph of the device after measurement (Fig. S7b), which shows shows dark spots in the upper left corner of the channel where prior to measurement the BN flake was wrinkled and poorly conformed to the gold surface. We expect that BN flakes without such mechanical defects will withstand a larger voltage before degrading.

## IV. WRINKLES IN BN FLAKES

The BN flakes in this study are transferred using a PMMA membrane[15]. This process leaves a layer of polymer residue all over the sample, even after cleaning in standard solvents. We remove this residue by annealing in oxygen at 500°C. One consequence of this is that the

15

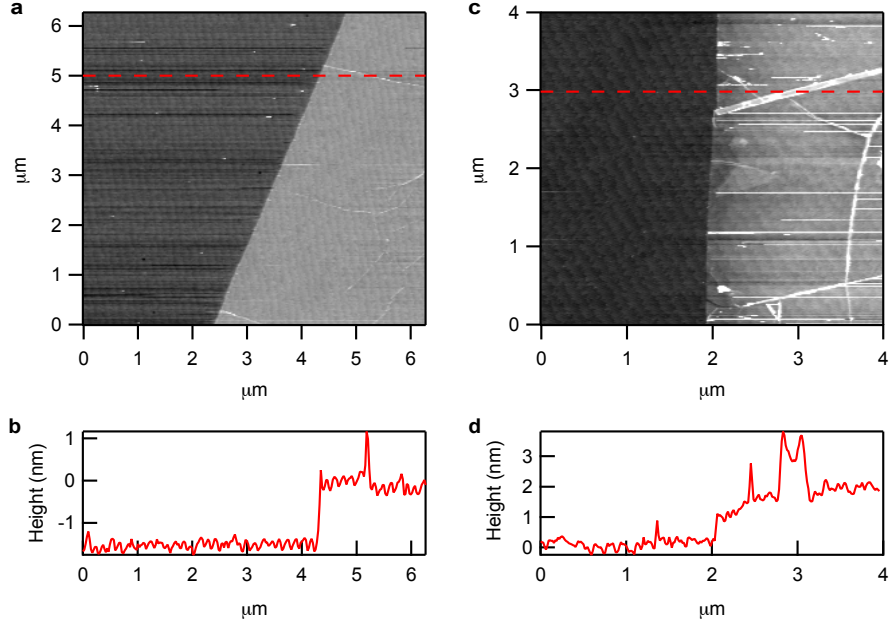

FIG. S8.  Wrinkles in BN flakes transferred to STO. **a**, Contact AFM topography of a flake showing minor wrinkles (right side of image) on STO (left), with typical wrinkle heights 0.5 to 1 nm. **b**, Height data for the scan along the red dashed line in **a**. **c**, Non-contact AFM topography of a much more severely wrinkled flake (right) on STO (left), where wrinkles are as high as 3 nm. The sample produced from this flake did not conduct between contacts, which were separated by thick wrinkles that spanned the channel. **d**, Height data for the scan along the red dashed line in **c**.

flakes tend to wrinkle, perhaps because of the different thermal expansion coefficients of STO and BN, or because small pockets of trapped contaminants escape from beneath the BN. If these wrinkles are only $\sim$ 1 nm tall (Fig. S8a,b), they present no obvious problem to our measurements: our highest mobility sample (Sample A) has small wrinkles, while Sample C has no wrinkles in the channel but still a lower mobility. However, some flakes have wrinkles several nm high (Fig. S8c,d); wrinkles this tall appear to prevent the underlying STO from being metallic at low temperature, just as one would expect based on electrostatic carrier accumulation.

The flakes as exfoliated are typically not wrinkled. The wrinkling is a byproduct of our particular transfer method and annealing process. Other methods, such as the dry transfer process based on polypropylene carbonate (PPC), are capable of wrinkle-free transfers[16].

16




\begin{thebibliography}{10}
\expandafter\ifx\csname url\endcsname\relax
  \def\url#1{\texttt{#1}}\fi
\expandafter\ifx\csname urlprefix\endcsname\relax\def\urlprefix{URL }\fi
\providecommand{\bibinfo}[2]{#2}
\providecommand{\eprint}[2][]{\url{#2}}

\bibitem{Fujimoto2013}
\bibinfo{author}{Fujimoto, T.} \& \bibinfo{author}{Awaga, K.}
\newblock \bibinfo{title}{{Electric-double-layer field-effect transistors with
  ionic liquids.}}
\newblock \emph{\bibinfo{journal}{Phys. Chem. Chem. Phys.}}
  \textbf{\bibinfo{volume}{15}}, \bibinfo{pages}{8983--9006}
  (\bibinfo{year}{2013}).

\bibitem{Schladt2013}
\bibinfo{author}{Schladt, T.~D.} \emph{et~al.}
\newblock \bibinfo{title}{{Crystal-Facet-Dependent Metallization in
  Electrolyte-Gated Rutile TiO$_2$ Single Crystals}}.
\newblock \emph{\bibinfo{journal}{ACS Nano}} \textbf{\bibinfo{volume}{7}},
  \bibinfo{pages}{8074--8081} (\bibinfo{year}{2013}).

\bibitem{Jeong2013}
\bibinfo{author}{Jeong, J.} \emph{et~al.}
\newblock \bibinfo{title}{{Suppression of metal-insulator transition in VO$_2$
  by electric field-induced oxygen vacancy formation.}}
\newblock \emph{\bibinfo{journal}{Science}} \textbf{\bibinfo{volume}{339}},
  \bibinfo{pages}{1402--1405} (\bibinfo{year}{2013}).

\bibitem{Li2013}
\bibinfo{author}{Li, M.} \emph{et~al.}
\newblock \bibinfo{title}{{Suppression of ionic liquid gate-induced
  metallization of SrTiO$_3$(001) by oxygen.}}
\newblock \emph{\bibinfo{journal}{Nano Lett.}} \textbf{\bibinfo{volume}{13}},
  \bibinfo{pages}{4675--4678} (\bibinfo{year}{2013}).

\bibitem{Petach2014}
\bibinfo{author}{Petach, T.~A.}, \bibinfo{author}{Lee, M.},
  \bibinfo{author}{Davis, R.~C.}, \bibinfo{author}{Mehta, A.} \&
  \bibinfo{author}{Goldhaber-Gordon, D.}
\newblock \bibinfo{title}{{Mechanism for the large conductance modulation in
  electrolyte-gated thin gold films}}.
\newblock \emph{\bibinfo{journal}{Phys. Rev. B}} \textbf{\bibinfo{volume}{90}},
  \bibinfo{pages}{081108} (\bibinfo{year}{2014}).

\bibitem{Ueno2008}
\bibinfo{author}{Ueno, K.} \emph{et~al.}
\newblock \bibinfo{title}{{Electric-field-induced superconductivity in an
  insulator.}}
\newblock \emph{\bibinfo{journal}{Nat. Mater.}} \textbf{\bibinfo{volume}{7}},
  \bibinfo{pages}{855--858} (\bibinfo{year}{2008}).

\bibitem{Ueno2011}
\bibinfo{author}{Ueno, K.} \emph{et~al.}
\newblock \bibinfo{title}{{Discovery of superconductivity in KTaO$_3$ by
  electrostatic carrier doping.}}
\newblock \emph{\bibinfo{journal}{Nat. Nanotech.}}
  \textbf{\bibinfo{volume}{6}}, \bibinfo{pages}{408--412}
  (\bibinfo{year}{2011}).

\bibitem{Ye2012}
\bibinfo{author}{Ye, J.~T.} \emph{et~al.}
\newblock \bibinfo{title}{{Superconducting dome in a gate-tuned band
  insulator.}}
\newblock \emph{\bibinfo{journal}{Science}} \textbf{\bibinfo{volume}{338}},
  \bibinfo{pages}{1193--1196} (\bibinfo{year}{2012}).

\bibitem{Ahn2003}
\bibinfo{author}{Ahn, C.~H.}, \bibinfo{author}{Triscone, J.-M.} \&
  \bibinfo{author}{Mannhart, J.}
\newblock \bibinfo{title}{{Electric field effect in correlated oxide systems.}}
\newblock \emph{\bibinfo{journal}{Nature}} \textbf{\bibinfo{volume}{424}},
  \bibinfo{pages}{1015--1018} (\bibinfo{year}{2003}).

\bibitem{Ahn1999}
\bibinfo{author}{Ahn, C.~H.} \emph{et~al.}
\newblock \bibinfo{title}{{Electrostatic Modulation of Superconductivity in
  Ultrathin GdBa$_2$Cu$_3$O$_{7-x}$ Films}}.
\newblock \emph{\bibinfo{journal}{Science}} \textbf{\bibinfo{volume}{284}},
  \bibinfo{pages}{1152--1155} (\bibinfo{year}{1999}).

\bibitem{Takahashi2006}
\bibinfo{author}{Takahashi, K.~S.} \emph{et~al.}
\newblock \bibinfo{title}{{Local switching of two-dimensional superconductivity
  using the ferroelectric field effect.}}
\newblock \emph{\bibinfo{journal}{Nature}} \textbf{\bibinfo{volume}{441}},
  \bibinfo{pages}{195--198} (\bibinfo{year}{2006}).

\bibitem{Boucherit2014}
\bibinfo{author}{Boucherit, M.} \emph{et~al.}
\newblock \bibinfo{title}{{Modulation of over 10$^{14}$ cm$^{-2}$ electrons in
  SrTiO$_3$/GdTiO$_3$ heterostructures}}.
\newblock \emph{\bibinfo{journal}{App. Phys. Lett.}}
  \textbf{\bibinfo{volume}{104}}, \bibinfo{pages}{182904}
  (\bibinfo{year}{2014}).

\bibitem{Yuan2009}
\bibinfo{author}{Yuan, H.} \emph{et~al.}
\newblock \bibinfo{title}{{High-Density Carrier Accumulation in ZnO
  Field-Effect Transistors Gated by Electric Double Layers of Ionic Liquids}}.
\newblock \emph{\bibinfo{journal}{Adv. Func. Mater.}}
  \textbf{\bibinfo{volume}{19}}, \bibinfo{pages}{1046--1053}
  (\bibinfo{year}{2009}).

\bibitem{Zhang2014}
\bibinfo{author}{Zhang, Y.~J.}, \bibinfo{author}{Oka, T.},
  \bibinfo{author}{Suzuki, R.}, \bibinfo{author}{Ye, J.~T.} \&
  \bibinfo{author}{Iwasa, Y.}
\newblock \bibinfo{title}{{Electrically switchable chiral light-emitting
  transistor.}}
\newblock \emph{\bibinfo{journal}{Science}} \textbf{\bibinfo{volume}{344}},
  \bibinfo{pages}{725--728} (\bibinfo{year}{2014}).

\bibitem{Ueno2010}
\bibinfo{author}{Ueno, K.}, \bibinfo{author}{Shimotani, H.},
  \bibinfo{author}{Iwasa, Y.} \& \bibinfo{author}{Kawasaki, M.}
\newblock \bibinfo{title}{{Electrostatic charge accumulation versus
  electrochemical doping in SrTiO$_3$ electric double layer transistors}}.
\newblock \emph{\bibinfo{journal}{App. Phys. Lett.}}
  \textbf{\bibinfo{volume}{96}}, \bibinfo{pages}{252107}
  (\bibinfo{year}{2010}).

\bibitem{YLee2011}
\bibinfo{author}{Lee, Y.} \emph{et~al.}
\newblock \bibinfo{title}{{Phase Diagram of Electrostatically Doped
  SrTiO$_{3}$}}.
\newblock \emph{\bibinfo{journal}{Phys. Rev. Lett.}}
  \textbf{\bibinfo{volume}{106}}, \bibinfo{pages}{136809}
  (\bibinfo{year}{2011}).

\bibitem{MLee2011}
\bibinfo{author}{Lee, M.}, \bibinfo{author}{Williams, J.~R.},
  \bibinfo{author}{Zhang, S.}, \bibinfo{author}{Frisbie, C.~D.} \&
  \bibinfo{author}{Goldhaber-Gordon, D.}
\newblock \bibinfo{title}{{Electrolyte Gate-Controlled Kondo Effect in
  SrTiO$_{3}$}}.
\newblock \emph{\bibinfo{journal}{Phys. Rev. Lett.}}
  \textbf{\bibinfo{volume}{107}}, \bibinfo{pages}{256601}
  (\bibinfo{year}{2011}).

\bibitem{Li2012}
\bibinfo{author}{Li, M.}, \bibinfo{author}{Graf, T.}, \bibinfo{author}{Schladt,
  T.~D.}, \bibinfo{author}{Jiang, X.} \& \bibinfo{author}{Parkin, S. S.~P.}
\newblock \bibinfo{title}{{Role of Percolation in the Conductance of
  Electrolyte-Gated SrTiO$_{3}$}}.
\newblock \emph{\bibinfo{journal}{Phys. Rev. Lett.}}
  \textbf{\bibinfo{volume}{109}}, \bibinfo{pages}{196803}
  (\bibinfo{year}{2012}).

\bibitem{Stanwyck2013}
\bibinfo{author}{Stanwyck, S.~W.}, \bibinfo{author}{Gallagher, P.},
  \bibinfo{author}{Williams, J.~R.} \& \bibinfo{author}{Goldhaber-Gordon, D.}
\newblock \bibinfo{title}{{Universal conductance fluctuations in
  electrolyte-gated SrTiO$_3$ nanostructures}}.
\newblock \emph{\bibinfo{journal}{App. Phys. Lett.}}
  \textbf{\bibinfo{volume}{103}}, \bibinfo{pages}{213504}
  (\bibinfo{year}{2013}).

\bibitem{Ueno2014}
\bibinfo{author}{Ueno, K.} \emph{et~al.}
\newblock \bibinfo{title}{{Effective thickness of two-dimensional
  superconductivity in a tunable triangular quantum well of SrTiO$_3$}}.
\newblock \emph{\bibinfo{journal}{Phys. Rev. B}} \textbf{\bibinfo{volume}{89}},
  \bibinfo{pages}{020508} (\bibinfo{year}{2014}).

\bibitem{Huijben2013}
\bibinfo{author}{Huijben, M.} \emph{et~al.}
\newblock \bibinfo{title}{{Defect Engineering in Oxide Heterostructures by
  Enhanced Oxygen Surface Exchange}}.
\newblock \emph{\bibinfo{journal}{Adv. Func. Mater.}}
  \textbf{\bibinfo{volume}{23}}, \bibinfo{pages}{5240--5248}
  (\bibinfo{year}{2013}).

\bibitem{Joshua2012}
\bibinfo{author}{Joshua, A.}, \bibinfo{author}{Pecker, S.},
  \bibinfo{author}{Ruhman, J.}, \bibinfo{author}{Altman, E.} \&
  \bibinfo{author}{Ilani, S.}
\newblock \bibinfo{title}{{A universal critical density underlying the physics
  of electrons at the LaAlO$_3$/SrTiO$_3$ interface.}}
\newblock \emph{\bibinfo{journal}{Nat. Commun.}} \textbf{\bibinfo{volume}{3}},
  \bibinfo{pages}{1129} (\bibinfo{year}{2012}).

\bibitem{BenShalom2010}
\bibinfo{author}{{Ben Shalom}, M.}, \bibinfo{author}{Ron, A.},
  \bibinfo{author}{Palevski, A.} \& \bibinfo{author}{Dagan, Y.}
\newblock \bibinfo{title}{{Shubnikov-De Haas Oscillations in
  SrTiO$_{3}$/LaAlO$_{3}$ Interface}}.
\newblock \emph{\bibinfo{journal}{Phys. Rev. Lett.}}
  \textbf{\bibinfo{volume}{105}}, \bibinfo{pages}{206401}
  (\bibinfo{year}{2010}).

\bibitem{Parish2003}
\bibinfo{author}{Parish, M.} \& \bibinfo{author}{Littlewood, P.}
\newblock \bibinfo{title}{{Non-saturating magnetoresistance in heavily
  disordered semiconductors}}.
\newblock \emph{\bibinfo{journal}{Nature}} \textbf{\bibinfo{volume}{426}},
  \bibinfo{pages}{1--4} (\bibinfo{year}{2003}).

\bibitem{Kozlova2012}
\bibinfo{author}{Kozlova, N.~V.} \emph{et~al.}
\newblock \bibinfo{title}{{Linear magnetoresistance due to multiple-electron
  scattering by low-mobility islands in an inhomogeneous conductor.}}
\newblock \emph{\bibinfo{journal}{Nat. Commun.}} \textbf{\bibinfo{volume}{3}},
  \bibinfo{pages}{1097} (\bibinfo{year}{2012}).

\bibitem{Joshua2013}
\bibinfo{author}{Joshua, A.}, \bibinfo{author}{Ruhman, J.},
  \bibinfo{author}{Pecker, S.}, \bibinfo{author}{Altman, E.} \&
  \bibinfo{author}{Ilani, S.}
\newblock \bibinfo{title}{{Gate-tunable polarized phase of two-dimensional
  electrons at the LaAlO$_3$/SrTiO$_3$ interface}}.
\newblock \emph{\bibinfo{journal}{PNAS}} \textbf{\bibinfo{volume}{110}},
  \bibinfo{pages}{9633--9638} (\bibinfo{year}{2013}).

\bibitem{Xie2013}
\bibinfo{author}{Xie, Y.}, \bibinfo{author}{Bell, C.}, \bibinfo{author}{Hikita,
  Y.}, \bibinfo{author}{Harashima, S.} \& \bibinfo{author}{Hwang, H.~Y.}
\newblock \bibinfo{title}{{Enhancing electron mobility at the
  LaAlO$_3$/SrTiO$_3$ interface by surface control.}}
\newblock \emph{\bibinfo{journal}{Adv. Mater.}} \textbf{\bibinfo{volume}{25}},
  \bibinfo{pages}{4735--4738} (\bibinfo{year}{2013}).

\bibitem{McCollam2014}
\bibinfo{author}{McCollam, A.} \emph{et~al.}
\newblock \bibinfo{title}{{Quantum oscillations and subband properties of the
  two-dimensional electron gas at the LaAlO$_3$/SrTiO$_3$ interface}}.
\newblock \emph{\bibinfo{journal}{APL Mater.}} \textbf{\bibinfo{volume}{2}},
  \bibinfo{pages}{022102} (\bibinfo{year}{2014}).

\bibitem{Xie2014}
\bibinfo{author}{Xie, Y.} \emph{et~al.}
\newblock \bibinfo{title}{{Quantum longitudinal and Hall transport at the
  LaAlO$_3$/SrTiO$_3$ interface at low electron densities}}.
\newblock \emph{\bibinfo{journal}{Solid State Commun.}}
  \textbf{\bibinfo{volume}{197}}, \bibinfo{pages}{25--29}
  (\bibinfo{year}{2014}).

\bibitem{Waqar2007}
\bibinfo{author}{Waqar, Z.}
\newblock \bibinfo{title}{{Hydrogen accumulation in graphite and etching of
  graphite on hydrogen desorption}}.
\newblock \emph{\bibinfo{journal}{J. Mater. Sci.}}
  \textbf{\bibinfo{volume}{42}}, \bibinfo{pages}{1169--1176}
  (\bibinfo{year}{2007}).

\bibitem{Bunch2008}
\bibinfo{author}{Bunch, J.~S.} \emph{et~al.}
\newblock \bibinfo{title}{{Impermeable atomic membranes from graphene sheets.}}
\newblock \emph{\bibinfo{journal}{Nano Lett.}} \textbf{\bibinfo{volume}{8}},
  \bibinfo{pages}{2458--2462} (\bibinfo{year}{2008}).

\bibitem{Das2013}
\bibinfo{author}{Das, D.}, \bibinfo{author}{Kim, S.}, \bibinfo{author}{Lee,
  K.-R.} \& \bibinfo{author}{Singh, A.~K.}
\newblock \bibinfo{title}{{Li diffusion through doped and defected graphene.}}
\newblock \emph{\bibinfo{journal}{Phys. Chem. Chem. Phys.}}
  \textbf{\bibinfo{volume}{15}}, \bibinfo{pages}{15128--15134}
  (\bibinfo{year}{2013}).

\bibitem{Ohno2011}
\bibinfo{author}{Ohno, H.}
\newblock \emph{\bibinfo{title}{Electrochemical Aspects of Ionic Liquids}}
  (\bibinfo{publisher}{Wiley}, \bibinfo{year}{2011}).

\bibitem{Rakhmilevitch2013}
\bibinfo{author}{Rakhmilevitch, D.} \emph{et~al.}
\newblock \bibinfo{title}{{Anomalous response to gate voltage application in
  mesoscopic LaAlO$_{3}$/SrTiO$_{3}$ devices}}.
\newblock \emph{\bibinfo{journal}{Phys. Rev. B}} \textbf{\bibinfo{volume}{87}},
  \bibinfo{pages}{125409} (\bibinfo{year}{2013}).

\bibitem{Connell2012}
\bibinfo{author}{Connell, J.~G.}, \bibinfo{author}{Isaac, B.~J.},
  \bibinfo{author}{Ekanayake, G.~B.}, \bibinfo{author}{Strachan, D.~R.} \&
  \bibinfo{author}{Seo, S. S.~A.}
\newblock \bibinfo{title}{{Preparation of atomically flat SrTiO$_3$ surfaces
  using a deionized-water leaching and thermal annealing procedure}}.
\newblock \emph{\bibinfo{journal}{App. Phys. Lett.}}
  \textbf{\bibinfo{volume}{101}}, \bibinfo{pages}{251607}
  (\bibinfo{year}{2012}).

\bibitem{Amet2013}
\bibinfo{author}{Amet, F.}, \bibinfo{author}{Williams, J.~R.},
  \bibinfo{author}{Watanabe, K.}, \bibinfo{author}{Taniguchi, T.} \&
  \bibinfo{author}{Goldhaber-Gordon, D.}
\newblock \bibinfo{title}{{Insulating Behavior at the Neutrality Point in
  Single-Layer Graphene}}.
\newblock \emph{\bibinfo{journal}{Phys. Rev. Lett.}}
  \textbf{\bibinfo{volume}{110}}, \bibinfo{pages}{216601}
  (\bibinfo{year}{2013}).

\end{thebibliography}
\end{document}